\pdfoutput=1 
\documentclass[prl,twocolumn,preprintnumbers,superscriptaddress,showpacs]{revtex4}

\usepackage{graphicx}
\usepackage{bm}
\usepackage{amsmath}
\usepackage{amssymb}
\usepackage[colorlinks,pdfstartview=FitH]{hyperref}
\hypersetup{linkcolor=blue,citecolor=blue,filecolor=black,urlcolor=blue}
\newcommand\sect[1]{\emph{#1.}---}

\newcommand\p{{\bm p}}
\newcommand\bP{{\bm P}}

\newcommand\hp{\bm {\hat p}}
\newcommand\abp{|\bm p|}
\newcommand\x{\bm {x}}
\newcommand\pd{\partial}
\newcommand\A{\bm {A}}
\newcommand\B{\bm {B}}

\newcommand\cO{{\cal O}}
\renewcommand\a{\bm {a}}
\renewcommand\b{\bm {b}}
\newcommand\dpp{{\frac{d^3\p}{(2\pi)^3}}}
\newcommand\letter{Letter}
\newcommand\cE{{\cal E}}
\newcommand\cI{{\cal I}}
\newcommand\bbeta{{\bm\beta}}

\begin{document}

\preprint{EFI-14-7}\preprint{RBRC-1068}
\author{Jing-Yuan~Chen}
\affiliation{Kadanoff Center for Theoretical Physics, University of 
Chicago, Chicago, Illinois 60637, USA}
\author{Dam T.~Son}
\affiliation{Kadanoff Center for Theoretical Physics, University of 
Chicago, Chicago, Illinois 60637, USA}
\author{Mikhail A.~Stephanov} 
\affiliation{Physics Department, University of Illinois at Chicago, Chicago, 
Illinois 60607, USA}
\affiliation{Kadanoff Center for Theoretical Physics, University of 
Chicago, Chicago, Illinois 60637, USA}
\author{Ho-Ung Yee}
\affiliation{Physics Department, University of Illinois at Chicago, Chicago, 
Illinois 60607, USA}
\affiliation{RIKEN-BNL Research Center, Brookhaven National Laboratory,
Upton, New York 11973-5000, USA}
\author{Yi Yin}
\affiliation{Physics Department, University of Illinois at Chicago, Chicago, 
Illinois 60607, USA}

\title{Lorentz Invariance in Chiral Kinetic Theory}
\date{April 2014}

\begin{abstract}

  We show that Lorentz invariance is realized nontrivially in the
  classical action of a massless spin-$\frac12$ particle 
  with definite helicity. We find that the
  ordinary Lorentz transformation is modified by a shift orthogonal to
  the boost vector and the particle momentum. The shift ensures
  angular momentum conservation in particle collisions and implies
  a nonlocality of the collision term in the Lorentz-invariant kinetic theory
  due to side jumps. 
  We show that 2/3 of the chiral-vortical
  effect for a uniformly rotating particle distribution can be attributed to
  the magnetic moment coupling required by the Lorentz invariance. We
  also show how the classical action can be obtained by taking
  the classical limit of the path integral for a Weyl particle.

\end{abstract}
\maketitle 

\sect{Introduction}%
The parity-odd response of a chiral medium and
its deep relationship to topology and quantum anomalies have attracted
significant theoretical interest.  Two such phenomena,
the chiral
magnetic and chiral vortical effects (CME and CVE), which is the appearance
of nonzero current in a magnetic field or when the system is in rotation, 
have been considered some time ago in astrophysical context
\cite{PhysRevD.20.1807,PhysRevD.22.3080}. 
More recently, the interest in such phenomena was
rekindled by developments in various subfields of physics.  
It was observed that charge-dependent
correlations can be used to detect the CME in heavy-ion collisions
\cite{Fukushima:2008xe}.  Independently, the chiral vortical effect has been 
found in a calculation using gauge/gravity duality 
\cite{Erdmenger:2008rm,Banerjee:2008th}, and 
a general argument based on second
law of thermodynamics was put forward in Ref.~\cite{Son:2009tf} to
demonstrate the generality of this result. 
The recent experimental discovery of ``3D
graphene'' \cite{2013arXiv1309.7892N,2014Sci...343..864L} 
brings closer the possibility of realizing the
materials with non-trivial chiral properties, such as Weyl
semimetals~\cite{2013arXiv1301.0330T}.

One promising approach to explore anomaly-related phenomena is the
kinetic theory, which can go beyond the regime of thermodynamic
equilibrium.  This kinetic approach is applicable when the external
fields and the interactions between the (quasi-)particles are
sufficiently weak, so each particle can be considered as moving along
a classical trajectory, punctuated by rare collisions.  Between
collisions, one has essentially a single-particle problem.  The
information about the quantum anomaly is encoded in the momentum-space
Berry curvature~\cite{Son:2012wh}.  The classical action for such a
motion can be derived either from a single-particle quantum
Hamiltonian
\cite{Stephanov:2012ki} or, more directly, from field theory \cite{Son:2012zy}.

There is, however, a puzzling aspect of the kinetic theory: it does
not have a manifest Lorentz symmetry, which it should inherit from the
original quantum field theory.  The issue was first raised in
Ref.~\cite{Son:2012zy}.  In this \letter\ we confirm the suggestion
made in Ref.~\cite{Son:2012zy} that Lorentz symmetry requires an
additional magnetic moment coupling term in the classical action of
the particle.  Unexpectedly, we also find that the Lorentz
transformation laws of the coordinates and momenta contain extra terms 
associated with particle spin.  Another
nontrivial consequence of the analysis is the appearance of a magnetization
current contribution to the total current, which is
required to reproduce the correct magnitude of the CVE.

\sect{Classical action.}%
We shall argue that the motion of a massless right-handed spin-$\frac12$
particle in an external electromagnetic field
is described, in the classical regime, by the following phase-space action,
\begin{equation}
  \label{eq:action}
    {\cI} = \int (\bm p + \bm A)\bm\cdot d\bm x - (\cE+\Phi) dt 
- \bm a_{\bm p}\bm\cdot d\bm p ,
\end{equation}
where $\a_\p$ is the Berry connection such that
\begin{equation}
  \label{eq:ba}
  \b \equiv \bm\nabla\times\a = \frac{\hp}{2\abp^2}, \qquad
  \hp \equiv \frac{\bm p}{|\bm p|}\,,
\end{equation}
while the dispersion relation~\cite{Son:2012zy}
\begin{equation}
  \label{eq:E}
  \cE\equiv |\bm p | -   \dfrac{\bm{\hat p \cdot B}}{2|\bm p|}
\end{equation}
is modified to linear order in the field by the magnetic moment
coupling.  (To describe a left-handed particle, one needs to flip the
sign of $\a$.)  Although we work in the convenient units
$\hbar=c=1$, it is easy to see, by restoring $\hbar$, that both the
Berry connection term in Eq.~(\ref{eq:action}) and the magnetic
coupling term in Eq.~(\ref{eq:E}) are of order $\cO(\hbar)$. Later in
the \letter, we will derive the action~(\ref{eq:action}) from the Weyl
Hamiltonian by taking the classical limit of a path integral, but for now we take it as the starting point.

\sect{Lorentz invariance}%
To zeroth order in $\hbar$ the action, $\int (\p+\A)\cdot d\x - (\abp+\Phi) dt$,
which is the action of a spinless particle,
is invariant with respect to the infinitesimal Lorentz boost
\begin{multline}
  \label{eq:boost}
  \delta_{\bbeta} \x = \bbeta t;\quad
  \delta_{\bbeta} t = \bbeta \cdot \x;\quad
  \delta_{\bbeta} \p = \bbeta \abp;\\
\delta_{\bbeta} \B = \bbeta \times \bm E;\quad
\delta_{\bbeta} \bm E = -\bbeta \times \bm B;\quad
\end{multline}
The $\cO(\hbar)$ terms in~(\ref{eq:action}) are not invariant with
respect to this boost , and the action changes by
\begin{equation}
 \label{eq:delta-beta-I}
  \delta_\bbeta \cI = \int\left[
\frac{\bbeta\times\hp}{2\abp}(\dot\p-\bm E - \hp\times\B)
+
\frac{\B\!\cdot\hp}{2\abp}\bbeta\!\cdot\!(\dot\x - \hp)
\right]dt .
\end{equation}
However, noting that the two expressions in parentheses 
are the variations of the $\cO(\hbar^0)$ part of action with
respect to $\x$ and $\p$ respectively,
one can find a modified Lorentz transformation for $\x$ and $\p$
\begin{equation}
  \label{eq:lorentz-mod}
  \delta'_\bbeta \x = \bbeta t  + \frac{\bbeta\times\hp}{2\abp};
\qquad 
  \delta'_\bbeta \p = \bbeta \cE + \frac{\bbeta\times\hp}{2\abp}\times \B;
\end{equation}
under which the action is invariant up to order $\hbar$ inclusively:
$\delta'_\bbeta\cI=\cO(\hbar^2)$.  This agrees with a result found
in Ref.~\cite{Skagerstam:1992er} for $\B=0$.

Thus, the action~(\ref{eq:action}) has, in fact, a hidden Lorentz
invariance, under which the position and the momentum of the particle
transform in a nontrivial manner.  We now give a physical
interpretation of the modified Lorentz transformations.

\sect{Angular momentum and side jump}%
We will assume for simplicity that $\bm E = \bm B =0$. 
Since the Berry connection comes into play when the particle
changes its momentum, we consider an
elastic scattering of two particles.
For simplicity, consider the process in the center of mass frame, and
assume zero impact parameter. The angular momentum conservation is
trivial in this frame: $\bm J_{\rm in}=\bm J_{\rm out}=0$ with both
orbital $\bm L$ and spin $\bm S$  contributions vanishing before and
after the collision. 

Let us now perform a Lorentz boost along the the direction of motion of one
of the incoming particles. Then the total angular momentum of incoming
particles is still zero $\bm J_{\rm in}=0$. However, the spins of the
outgoing particles no longer cancel each other, since their
momenta are not collinear in the new frame. That means that the
orbital momentum of the outgoing pair should be nonzero, which would be
impossible if the particle trajectories were going through a single
collision point. 

\begin{figure}
  \centering
  \includegraphics[width=12em]{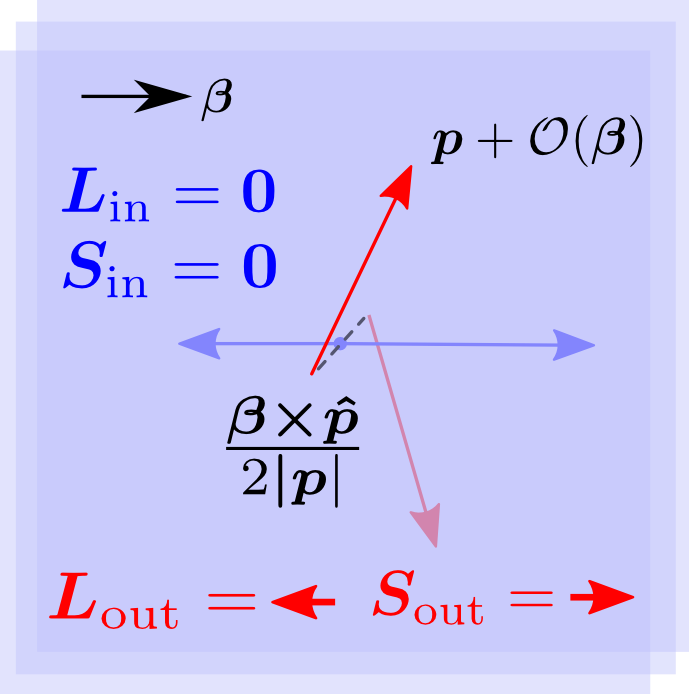}
  \caption{(color online) Side jump}
  \label{fig:side-jump}
\end{figure}

However, the modified Lorentz transformation in
Eq.~(\ref{eq:lorentz-mod}) shifts the trajectory in the direction
perpendicular to the boost and the particle momentum: $\Delta \x = \bbeta\times\hp/(2\abp)$. Since the
momenta of the particles, $\p$ and $-\p$, are opposite before the
boost, the shifts are also opposite. As a result the two outgoing
particles are moving in two parallel planes. It is easy to check that
such a shift leads to a contribution to the orbital momentum
\begin{equation}
  \label{eq:Lout}
  \bm L_{\rm out} = \frac{\bbeta\times \hp}{\abp}\times\p
\end{equation}
equal and opposite to the total spin of the outgoing particles
\begin{equation}
  \label{eq:Sout}
  \bm S_{\rm out} = \delta_\bbeta (\hp) = \frac{\bbeta 
  - \hp(\bbeta\cdot\hp)}\abp
  = -\bm L_{\rm out}.
\end{equation}

Therefore, collisions of two particles with spin involves a shift in
the position.  This is similar to the ``side jump'' phenomenon in
impurity scatterings with spin-orbit interaction~\cite{Berger:1970}.
The magnitude of the side
jump is frame-dependent and does not depend on the details of the collision.
This phenomenon has a classical analog: the center
of mass of a spinning extended particle is frame 
dependent~\cite{Costa:2011zn}.
We expect the side jump to be important for
constructing Lorentz invariant chiral kinetic theory with collisions,
and that in such a theory the collision kernel must be nonlocal in
space and time.

\sect{Lorentz algebra}%
We now check that the modified Lorentz transformations satisfy the algebra
of the Lorentz group.
It is well-known that the commutator of the ordinary Lorentz
transformations is a rotation. For example,
$  
[\delta_{\bbeta_1},\delta_{\bbeta_2}]\x = \bm\varphi\times\x,
$ where $
\bm \varphi \equiv  {\bbeta_1}\times  {\bbeta_2} .
$
For the modified Lorentz transformation, however,
\begin{equation}
  \label{eq:d1d2-mod}
  [\delta_{\bbeta_1}',\delta_{\bbeta_2}']\x = \bm\varphi\times\x
- \hp\,\frac{\bm\varphi\cdot\hp}{\abp}\,;\quad
[\delta_{\bbeta_1}',\delta_{\bbeta_2}']t =  - \frac{\bm\varphi\cdot\hp}{\abp}
 \,,
\end{equation}
where, for simplicity, we set the electromagnetic field to zero. We
see that the rotation is accompaniend by an additional shift $\delta
t=-\bm\varphi\cdot\hp/\abp$, $\delta\x=\delta t\, \hp$ which, by virtue
of the fact that $d\x=dt\,\hp$ on equations of motion, is an invariance
of the action (for a classical trajectory it amounts to time
reparametrization).

Similar results hold in the presence of the external electromagnetic field.

\sect{Chiral vortical effect}%
Another nontrivial consequence of the magnetic moment coupling is a
contribution to the current which turns out to be
essential for reproducing correct value of the chiral vortical effect.

The current is determined by variation of the
action with respect to external gauge potential $\A$. The resulting
single-particle current (in zero field) is given by
\begin{equation}
  \label{eq:J}
  \bm J(\x,t) \equiv \frac{\delta \cI}{\delta \A(\x,t)}\Bigg|_{\A=0}
\!= \left(\hp - \frac{\hp}{2\abp}\! \times\! \bm\nabla\right)\delta^3(\x-\x'(t)) 
\end{equation}
where $\x'(t)$ is the position of the particle at time $t$. Consider
now an ensemble of particles with a distribution function $f$.  The
corresponding current is given by
\begin{equation}
  \label{eq:Jf}
  \bm J(\x,t) = \int\! \dpp 
  \Bigl(\hp f - \frac\hp{2\abp}\times\bm\nabla\! f\Bigr).
\end{equation}
The first term is the classical Liouville current, while the second
term, which is due to the magnetic moment coupling, is
$\cO(\hbar)$.  It is trivially conserved because it can be written as $\bm\nabla\times \bm M$, where
\begin{equation}
  \label{eq:M}
  \bm M = \int\! \dpp \frac{\hp}{2\abp} f,
\end{equation}
is the total magnetization (the sum of the magnetic moments).  However,
this contribution is needed to make the current a Lorentz vector and, as
we shall now show, to reproduce the correct magnitude of the CVE.

Consider a distribution $f$ such that there exist a
frame in which the distribution is isotropic in momentum. Denoting the
energy of particles in this frame $\epsilon'$ we can write
$f=f(\epsilon')$. Now consider a distribution which, in addition, varies
very slowly in space because the velocity $\bm u$ of the frame in which the
distribution in momentum is isotropic varies very slowly with space
point $\x$. Since the distribution function is a Lorentz scalar we can write
the distribution in the lab frame as $f=f(\epsilon')$, where
$
\epsilon'=\epsilon-\p\cdot \bm u - \lambda\hp\cdot \bm \omega
$
is the energy in the locally comoving frame expressed in terms of the
lab energy $\epsilon$ and momentum $\p$ and the helicity of the
particle $\lambda=\frac12$. The last term is present if the velocity
distribution has vorticity $\bm\omega=\bm\nabla\times\bm u/2$ since the
particle carries intrinsic angular momentum $\lambda\hp$.

The shift $-\lambda\hp\cdot\bm\omega$ arises naturally when $f$ is a
local equilibrium solution of Boltzmann equation.
The detailed balance dictates 
that, for fermions, $\ln[f/(1-f)]$ is a linear function of the
conserved quantities $\epsilon$, $\p$ and angular momentum $\bm j$,
i.e., $-\beta(\epsilon- \p\cdot\bm u_0-\bm j \cdot \bm\alpha)$ with
some constants  $\beta$, $\bm u_0$ and $\bm \alpha$. Inserting $\bm
j=\x\times\p+\lambda\hp$ gives $-\beta(\epsilon- \p\cdot\bm u-\lambda \hp\cdot \bm \alpha)$
with $\bm u\equiv \bm u_0+\bm\alpha\times\x$. This means the equilibrium
distribution could be inertially moving as well as rotating and that $\bm\alpha=\bm\omega$.

Substituting the distribution 
$f(\epsilon-\p\!\cdot\! \bm u - \lambda\hp\!\cdot\! \bm
\omega)$ into Eq.~(\ref{eq:Jf}) and Taylor expanding to linear order
in $\bm u$ and $\bm\omega$ one finds that magnetization current
contributes 2/3 of the total current:
\begin{multline}
  \label{eq:Jp}
  \bm J = -\!\int\! \dpp\frac{\partial
f}{\partial\epsilon}\,\frac12\left[\,\hp (\hp\cdot\bm\omega) 
- \hp\times\bm\nabla(\hp\cdot\bm u)\,\right]\\
= - \frac{\bm\omega}2\! \int\!\dpp \frac{\partial
f}{\partial\epsilon} \left[\,\frac13+\frac23\,\right].
\end{multline}
where we used the isotropy of $f$ to replace $\hat p^i\hat p^j$ by
$\delta^{ij}/3$ under the integral. Now using $\epsilon=\abp$, taking
the integral over angular directions of $\p$ and then integrating by
parts, we find for the current:
\begin{equation}
  \label{eq:Jpp}
  \bm J = \frac{\bm \omega}{4\pi^2}\!\int_0^\infty\! d\epsilon\, 2\epsilon f,
\end{equation}
which agrees with the expression for the CVE obtained from the CME by the
substitution $\bm B \to 2\epsilon\bm\omega$ (for isotropic
distributions)~\cite{Stephanov:2012ki}.

\sect{Classical action from path integral}%
We now show that the action~(\ref{eq:action}), including the magnetic
moment coupling, can be derived systematically from the path
integral. This derivation is complementary to the
previously developed wave-packet
approach (see Ref.~\cite{Chang:2008zza} for a review). We
start from a path-integral representation of a transition amplitude
for the Weyl Hamiltonian in an external field
\begin{equation}
  \label{eq:H-field}
  {\cal H} = \bm \sigma\cdot (\p - \A(\x,t)) + \Phi(\x,t)
\end{equation}
where $\x$ and $\p$ are canonically conjugate operators of position
and momentum: $[x^i,\, p^j]=i\delta^{ij}$. Inserting the sums over
complete sets of momentum and coordinate eigenstates at
infinitesimally spaced points in time, the transition amplitude
from a given state $i$ to state $f$ can be rewritten as a matrix
element of the path-ordered products of $2\times2$ matrices:
\begin{multline}
  \label{eq:amplitude}
  {\cal A}_{fi}
=\langle f| {\cal P} e^{-i\!\int\! {\cal H} dt} |i\rangle
\\=
\int\! {\cal D}\p\,{\cal D}\x \left[ 
{\cal P}e^{i\! \int\!  
\p\cdot d\x  - \bm\sigma\cdot(\p-\A)dt - \Phi dt
} 
\right]_{fi}.
\end{multline}
Here, and from now on, the symbols $\p$ and $\x$ refer to ordinary
$c$-number integration variables. By using a $2\times2$ matrix 
$V_{\p}$ satisfying
$V_{\p}^\dag\bm\sigma\!\cdot\!\p V_{\p}=\sigma_3 \abp$, one can
diagonalize the matrix $\exp(-i\bm\sigma\!\cdot\!\p\,\Delta t)$ on each
infinitesimal interval of the trajectory and recast the amplitude as
  \begin{multline}
    \label{eq:A0}
    {\cal A}_{fi}=\Big[\lim_{\Delta t\to0}\int\! 
   \prod_t d\p_{(t)} d\x_{(t)} e^{i\p\cdot\Delta \x}
e^{-i\sigma_3\abp\Delta t-i\Phi\Delta t}
\\
V_{\p}^\dag e^{i\bm\sigma\cdot\A\Delta t}V_{\p-\Delta \p}
\Big]_{fi}
  \end{multline}
where $\Delta \p \equiv \p_{(t)}-\p_{(t-\Delta t)}\equiv \p-\p'$.

In the classical regime, we can neglect off-diagonal
elements of the propagator matrix and consider only the contribution to
the phase factor given by the diagonal matrix elements between
positive-energy eigenvectors of $\sigma_3\abp$ ($i=f=+$). In particular we need to
evaluate the factor
\begin{equation}\label{eq:VeV++}
\left[
V_{\p}^\dag e^{i\bm\sigma\cdot\A\Delta t}V_{\p-\Delta    \p}
\right]_{++}
= 
u_{\p}^\dag e^{i\bm\sigma\cdot\A\Delta t}u_{\p-\Delta \p}
,
\end{equation}
where $u_\p$ is the positive energy eigenvector---the solution of the Weyl
equation:  $\bm\sigma\cdot\p\, u_\p = \abp u_\p $. Using the Gordon
identity for $u^\dag_\p\bm\sigma u_{\p'}$ and iterating the
 identity to
linear order in $\Delta\p =\p-\p'$
we rewrite Eq.~(\ref{eq:VeV++}) as
\begin{multline}\label{eq:VeVaA}
\left[
V_{\p}^\dag e^{i\bm\sigma\cdot\A\Delta t}V_{\p-\Delta\p}
\right]_{++}
= u_\p^\dag u_{\p'}
 \exp\Bigg[
i\,\frac{\hp+\hp'}2 \cdot\A\Delta t 
\\
+ \,  \frac{\Delta\p\times\hp}{2\abp}\cdot\A
\Delta t\Bigg]+ \cO(\Delta\p^2,\Delta t^2).
\end{multline}
The first term in the square brackets combines with neighboring factors
$e^{-i\abp\Delta t}$ in the path-ordered product in Eq.~(\ref{eq:amplitude})
to replace $\abp$ with $|\p -\A|\approx \p - \hp\cdot \A +
\cO(\A^2/\abp)$ \footnote{Gauge invariance means that
  tracking $\cO(\A^2)$ terms would require us to consider
  contributions to the action at order $\B^2$, beyond the
  $\cO(\hbar)$ order we are working within. Note that
we can always choose $\A$ to vanish in a given space-time point and,
because $\B$ is small compared to the relevant scale $\p^2$, 
$\A$ will remain small compared to $\abp$ 
in a patch much larger than de Broglie wavelength $1/\abp$ 
around such a point.}.

Naively, we could neglect the second term in the square brackets in
Eq.~(\ref{eq:VeVaA}) because it contains an additional factor
$\Delta\p$. However, we need to keep in mind that $\p$ and $\p'$ are independent
integration variables and the difference $\Delta\p$ is not small in
general. Rather, it is the factor
$\prod\exp(i\p\cdot\Delta\x)=\prod\exp(-i\x\cdot\Delta \p)$ which, upon
integration over $\x$, makes rapidly oscillating contributions at large
$\Delta\p$ cancel out. It is easy to see that if $\Delta\p$
multiplies a function of $\x$ the result of integration is the same as
if we replaced $\Delta \p$ with $-i\pd/\pd\x$ as in this example:
\begin{equation}
  \label{eq:F}
  \int\! dx\, e^{-ix\Delta p} \Delta p F(x) 
  = -i\!\int\! dx\, e^{-ix\Delta p}\,  \frac{dF(x)}{dx}\,.
\end{equation}
This relation is the path-integral representation of the canonical
commutation relation between $x$ and $p$ (similar to the commutation
relation between coordinate and velocity discussed in 
Ref.~\cite{Feynman-thesis}). 
This means that we cannot consider $\Delta \p$ as small in the
second term in Eq.~(\ref{eq:VeVaA}) if $\A$ depends on $\x$. Replacing
$\Delta\p$ with  $-i\pd/\pd\x$ we find that this term contributes
$i\hp\cdot\B/(2\abp)\Delta t$ to the phase, and thus represents the
interaction energy of the particle's magnetic moment.

Finally, the factor  $u_\p^\dag u_{\p'}=\exp(-i\a_\p\!\cdot\!\Delta\p)$ is the Berry phase. If we
express it using the physical (gauge-invariant) momentum $\bm P =
\p-\A$, we can, to linear order in $\A$, write
\begin{multline}
  \label{eq:ap}
\langle \ldots u_\p^\dag u_{\p'} \ldots\rangle
= \langle\ldots   (1-i\a_\p\cdot\Delta\p) \ldots\rangle
\\
= \langle\ldots   (1-i(\a_\bP\cdot\Delta\bP 
+ \Delta\p\times\b\cdot\A)  \ldots\rangle
\\
= \langle\ldots   (1+\b\cdot\B -i\a_\bP\cdot\Delta\bP)  \ldots\rangle
\\
=  \langle\ldots  (1+\b\cdot \B)e^{-i\a_\bP\cdot\Delta\bP} \ldots\rangle
\end{multline}
where $\langle\ldots\,\ldots\rangle$ denote remaining factors and
limits in the path
integral in Eq.~(\ref{eq:A0}) and in the third line we replaced $\Delta\p$ with
$-i\pd/\pd\x$ as before. We find that if we change variables to
physical momentum $\bP$, the factor $u_\p^\dag u_{\p'}$, expanded  to
order $\Delta \bP$, and under path integration, cannot be treated as 
a pure phase. The magnitude factor $(1+\b\cdot\B)$ in
Eq.~(\ref{eq:ap}) combined with the
path-integral measure $d\x\, d\bP$ gives the correct conserved
(up to the anomaly~\cite{Son:2012wh,Stephanov:2012ki})
Liouville measure for a Weyl particle in the magnetic field.

\sect{Conclusions}%
We have shown that the theory of a single particle with spin-1/2 and
definite helicity can be made Lorentz-invariant if one includes one
term in the action that corresponds to the interaction
between the particle's magnetic moment with the magnetic field.  The
magnitude of the magnetic moment is completely determined by Lorentz
invariance.  We have also shown that the Lorentz transformations of
the particle's coordinates and momentum components are nontrivial, and
that they are related to the side jumps in scattering processes.

Although our action has Lorentz symmetry, it is not written in a
manifestly Lorentz invariant manner.  We are currently developing a
manifestly Lorentz-invariant formulation, which will be reported
elsewhere.  It would also be interesting to generalize this analysis to
higher dimensions and non-abelian anomalies \cite{Dwivedi:2013dea}.

From the equation of motion of a single particle one can go to the
kinetic description in terms of a Boltzmann equation.  We expect that
the side jumps required by Lorentz invariance are necessary for the
collision term in the Boltzmann equation to be consistent with angular
momentum conservation.  Understanding how to write down a correct
kinetic theory of chiral particles, including their interactions, will
provide a link, so far missing, between quantum field theory and
hydrodynamics with anomalies and would allow, in particular, treatments
of processes far from equilibrium in theories with anomalies.

The authors would like to thank D.~Kharzeev, E.~Martinec, M.~Stone,
and R.~Wald for discussions and the Simons Center for Geometry and
Physics at Stony Brook University for hospitality. This work is
supported, in part, by a Simon Investigator grant from the Simons
Foundation and the US DOE grants Nos.\ DE-FG0201ER41195 and
DE-FG02-13ER41958.

\bibliography{Lorentz}

\end{document}